\documentclass[twocolumn,prl,superscriptaddress]{revtex4-2}

\usepackage{amsmath,amssymb}
\usepackage{graphicx}
\usepackage{dcolumn}
\usepackage{bm}
\usepackage{xr}
\usepackage[margin=0.75in]{geometry}
\usepackage{physics}

\usepackage{hyperref}        
\hypersetup{colorlinks,breaklinks,
             linkcolor=black,urlcolor=black,
             anchorcolor=black,citecolor=black}

\newcommand{\tauR}{\tau_{\mathrm{r}}}

\makeatletter
\newcommand*{\addFileDependency}[1]{
  \typeout{(#1)}
  \@addtofilelist{#1}
  \IfFileExists{#1}{}{\typeout{No file #1.}}
}
\makeatother

\newcommand*{\myexternaldocument}[1]{%
    \externaldocument{#1}%
    \addFileDependency{#1.tex}%
    \addFileDependency{#1.aux}%
}

\myexternaldocument{supplementary}

\begin{document}

\title{Hunting for Maxwell's Demon in the Wild}

\author{Johan du Buisson}
\thanks{These authors contributed equally.}
\affiliation{Department of Physics, Simon Fraser University, Burnaby, BC, V5A 1S6, Canada.}

\author{Jannik Ehrich}
\thanks{These authors contributed equally.}
\affiliation{Department of Physics, Simon Fraser University, Burnaby, BC, V5A 1S6, Canada.}

\author{Matthew P.\ Leighton}
\thanks{These authors contributed equally.}
\affiliation{Department of Physics, Simon Fraser University, Burnaby, BC, V5A 1S6, Canada.}
\affiliation{Department of Physics and Quantitative Biology
Institute, Yale University, New Haven, CT, 06511, USA}

\author{Avijit Kundu}
\affiliation{Department of Physics, Simon Fraser University, Burnaby, BC, V5A 1S6, Canada.}

\author{Tushar K.\ Saha}
\affiliation{Department of Physics, Simon Fraser University, Burnaby, BC, V5A 1S6, Canada.}
\affiliation{Current Address: MKS Instruments, Inc., 130-13500 Verdun Place, Richmond, BC, V6V 1V2, Canada}

\author{John Bechhoefer}
\affiliation{Department of Physics, Simon Fraser University, Burnaby, BC, V5A 1S6, Canada.}

\author{David A.\ Sivak}%
\email{dsivak@sfu.ca}
\affiliation{Department of Physics, Simon Fraser University, Burnaby, BC, V5A 1S6, Canada.}

\begin{abstract}
The paradox of Maxwell's demon motivated the development of information thermodynamics and the creation of nanoscale information engines. We now understand that machines such as the molecular motors within cells can in principle harvest fluctuations and thereby operate as a Maxwell demon---but do they? Answering this question would seemingly require simultaneous measurement of all system degrees of freedom, which is generally intractable in single-molecule experiments. Here, we derive a simple statistical estimator to infer both the direction and magnitude of subsystem heat flows, and thus determine whether---and how strongly---a motor operates as a Maxwell demon. The estimator uses only trajectory measurements for a single degree of freedom. Simulating both colloidal information engines and kinesin molecular motors, we show that our estimator can precisely and accurately detect Maxwell-demon behavior with experimentally accessible resolution and quantities of data. Moreover, we find that kinesin transitions to a Maxwell-demon mechanism in the presence of nonequilibrium noise, with a corresponding increase in velocity consistent with experiments. These findings suggest that molecular motors may have evolved to leverage active fluctuations within cells. 
\end{abstract}

\maketitle

The thought experiment of Maxwell's demon, first proposed in 1867~\cite{Maxwell1867}, drove interest in clarifying the connection between information and thermodynamics. The quest to exorcize Maxwell's demon by showing overall consistency with the second law of thermodynamics led to the modern field of information thermodynamics, where it is now possible to directly quantify information processing in stochastic systems far from equilibrium~\cite{Parrondo2015_Thermodynamics}. In parallel, technological progress in the control of mesoscale systems has allowed for Maxwell's demon to be realized experimentally~\cite{goerlich2025experimental} in a variety of settings, including colloidal~\cite{Toyabe2010_Experimental,roldan2014universal,Paneru2018_Losless,Admon2018_Experimental,saha2021maximizing}, optical~\cite{thorn2008experimental,vidrighin2016photonic,kumar2018sorting}, electronic~\cite{Koski2014_Experimental_Realiz,Koski2015_On-chip,Chida2017_Power}, single-molecule~\cite{Ribezzi2019_Large,amano2022insights}, mechanical~\cite{barros2024probabilistic}, granular~\cite{lagoin2022human}, and active-particle~\cite{chor2023many} systems. These ``information engines" extract energy from the surrounding environment and can attain performance comparable to evolved molecular motors within biological organisms~\cite{du2024performance}.

Molecular motors such as kinesin fulfill a wide range of important functions within living cells, accomplishing their tasks by transducing free energy between different forms. These nanoscale machines feature stochastic dynamics and have energy scales comparable to the ambient thermal energy, $k_\mathrm{B}T$~\cite{Brown2019theory}. In such settings, stochastic fluctuations are omnipresent, and hence information is a relevant thermodynamic resource for these systems, meaning that they can in principle interconvert information and other forms of free energy~\cite{mcgrath2017biochemical,leighton2024flow}. This, in addition to the comparable performance of experimental information engines~\cite{du2024performance}, raises the question of whether biological molecular machines have evolved to use information as a thermodynamic resource. Thus, we seek to determine whether, and if so under what conditions, molecular motors can operate as Maxwell demons. Previous work has explored the related, but distinct, dichotomy between power-stroke and Brownian-ratchet dynamics for molecular motors~\cite{vale1990protein,wagoner2016molecular,penocchio2024power}; we focus on the purely thermodynamic phenomenon of Maxwell-demon behavior, characterized by heat flow into the system~\cite{leighton2024flow}.

Here we focus on transport motors, such as kinesin, that consume chemical energy to pull cargo along cytoskeletal filaments~\cite{howard2002mechanics}. These motors are conceptually similar to experimental implementations of information engines: they take discrete steps to transport cargo subject to thermal fluctuations and external forces. Recent experiments show that nonequilibrium fluctuations can increase the velocity of kinesin motors pulling cargo against opposing forces~\cite{ariga2021noise}, with similar behavior observed for information engines~\cite{Saha2022_Information}. Subsequent theoretical work revealed that an internal information flow, a hallmark of Maxwell-demon behavior, is required to produce net output power in the presence of different sources of fluctuations~\cite{leighton2024information}. Determining whether a given system behaves as a Maxwell demon requires measuring subsystem heat flows; such knowledge is generally not directly observable with today's experiments, which typically only resolve one degree of freedom in multicomponent systems. As such, we turn to thermodynamic inference: inferring thermodynamic quantities of interest from experimentally measurable observables~\cite{pietzonka2016universal,leighton2023inferring,di2024variance}.

In this Letter, we introduce a method for estimating subsystem heat flows in bipartite stochastic systems from measurements of mean squared displacement. Our main result, Eq.~\eqref{eq:heatestimator}, is a statistical estimator that can accurately and precisely estimate the heat flow with experimentally accessible quantities of data. We benchmark this new estimator on simulations parameterized by a well-characterized experimental realization of a Maxwell demon, demonstrating the ability to infer the heat flow and hence distinguish between conventional-engine and information-engine operational modes. We then apply the estimator to quantify heat flows in simulations of a kinesin motor pulling a diffusive cargo. The estimator successfully infers the direction of heat flow for experimentally accessible spatiotemporal resolutions and trajectory durations, and it detects a transition to Maxwell-demon behavior accompanying an increase in velocity when nonequilibrium noise is applied to the cargo.

\begin{figure}[t]
    \centering
    \includegraphics[width =\linewidth]{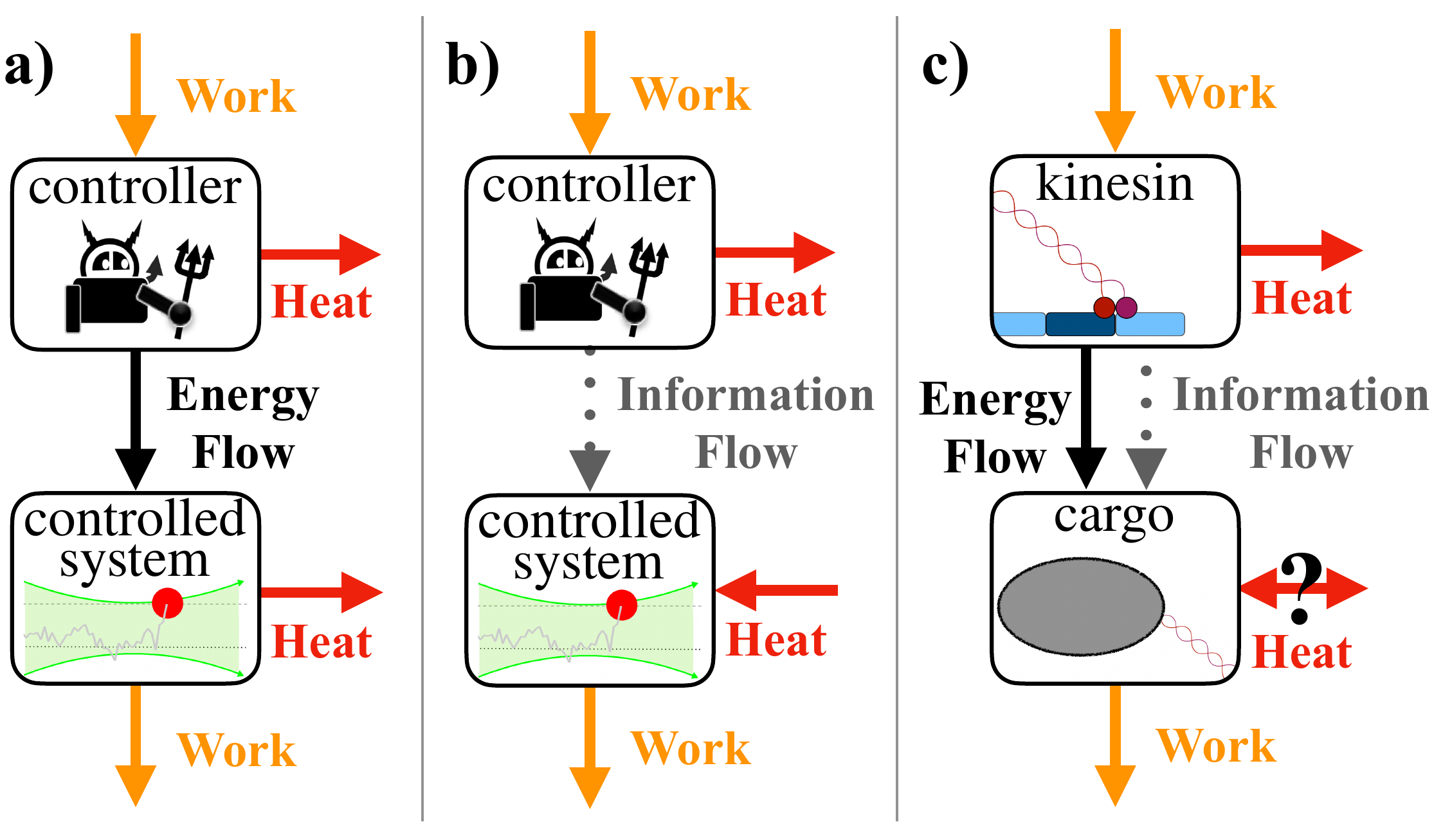}
    \caption{Bipartite thermodynamics of two model systems. a,b) An experimental realization of a Maxwell demon, which in opposite limits operates as either a conventional engine (a) or an information engine (b). c) Kinesin motor pulling a diffusive cargo. Information-engine schematics adapted from Ref.~\cite{saha2021maximizing} (copyright 2021 National Academy of Sciences).}
    \label{fig:fig1}
\end{figure}

\emph{Bipartite thermodynamics of Maxwell demons}---The thermodynamics of Maxwell demons are naturally considered through the lens of bipartite thermodynamics, where a system is decomposed into two thermodynamic subsystems (Fig.~\ref{fig:fig1}). For a kinesin-cargo system (Fig.~\ref{fig:fig1}c), the subsystems are the kinesin motor ($Y$) and cargo ($X$). Information engines (Fig.~\ref{fig:fig1}b) can similarly be decomposed into a controller and controlled system. As reviewed in Ref.~\cite{Ehrich2023_Energy}, each subsystem can exchange energy with its environment in the form of work and heat, and the subsystems can exchange free energy through energy and information flows. Each subsystem satisfies a first law describing the local balance of energy.

At steady state, the second law forbids the total average heat flow $\dot{Q}$ from being positive (under the convention that heat flow \textit{into} the system is positive), so that the entropy production rate $\dot{\Sigma}$ is nonnegative, $\dot{\Sigma} = -\beta\dot{Q} \geq 0$. The bipartite structure permits decomposition of both the heat flow and entropy production rate into respective contributions from the $Y$ and $X$ subsystems: $\dot{\Sigma} = \dot{\Sigma}_Y + \dot{\Sigma}_X$ and $\dot{Q} = \dot{Q}_Y + \dot{Q}_X$. As both $Y$ and $X$ subsystems are themselves valid thermodynamic systems, each satisfies a local second law, $\dot{\Sigma}_{Y/X}\geq 0$.

One might expect that the second law holds for each subsystem in the same way as for the system as a whole, so that both $\dot{Q}_Y<0$ and $\dot{Q}_X<0$. Surprisingly, this is not the case: one subsystem (say, $X$) can have a positive heat flow from the environment, $\dot{Q}_X>0$. This possibility, which seemingly violates the second law of thermodynamics, is the crux of the Maxwell-demon paradox. The paradox is resolved by considering the mutual information $I[X;Y]$ between $Y$ and $X$~\cite{Cover2006_Elements}, a thermodynamic resource that must be accounted for~\cite{Parrondo2015_Thermodynamics}. Both subsystems can change the mutual information at rates $\dot{I}_Y$ and $\dot{I}_X$, known as \emph{information flows}~\cite{horowitz2014thermodynamics}. The information flow $\dot{I}_Y$ is the rate at which $Y$ (through its dynamics) changes the mutual information, which is then a thermodynamic resource that $X$ can use to extract energy from thermal fluctuations~\cite{leighton2024flow}. Because all quantities are constant at steady state, the total time derivative of the mutual information is zero, hence $0 = \mathrm{d}_tI = \dot{I}_Y + \dot{I}_X$.

Information flows allow for a thermodynamically consistent formulation of subsystem-level second laws:
\begin{subequations}
\begin{align}\label{eq:ysecondlaw}
\dot{\Sigma}_Y & = -\beta\dot{Q}_Y - \dot{I}_Y \geq 0\ ,\\
\label{eq:xsecondlaw}
\dot{\Sigma}_X & = -\beta\dot{Q}_X + \dot{I}_Y \geq 0\ .
\end{align}
\end{subequations}
From Eq.~\eqref{eq:xsecondlaw}, an apparent local violation of the second law ($\dot{Q}_X>0$) is possible, provided the system supports a sufficiently large information flow, $\dot{I}_Y\geq \beta\dot{Q}_X$. To satisfy Eq.~\eqref{eq:ysecondlaw}, this information flow in turn requires that the heat flow from the $Y$ subsystem satisfy $-\beta\dot{Q}_Y \geq\dot{I}_Y$. Thus, $-\beta\dot{Q}_Y\geq \beta\dot{Q}_X$, satisfying the global second law.

Although Maxwell demons have had many definitions~\cite{sanchez2019nonequilibrium,ciliberto2020autonomous,freitas2021characterizing}, all share a common feature: some subsystem appears to locally break the second law unless information is properly accounted for. The subsystem second laws~\eqref{eq:ysecondlaw} and \eqref{eq:xsecondlaw} permit quantitative definition of Maxwell-demon behavior: one of the subsystem heat flows is positive. Without loss of generality, let $X$ be the subsystem with $\dot{Q}_X > 0$. Since measuring the heat flow $\dot{Q}_X$ suffices to detect a Maxwell demon, in the next section we derive a method for estimating $\dot{Q}_X$ directly from experimentally accessible measurements.

\emph{Subsystem heat-flow estimator}---We develop our framework for a transport motor or controller $Y$ pulling a diffusive cargo $X$ in one dimension. At this point we make no assumptions as to the details of the motor dynamics. We assume the cargo dynamics obey an overdamped Langevin equation, with a linear coupling force:
\begin{equation}
\label{eq:xdynamics}
    \dot{x} = \gamma^{-1} \left[\kappa(y-x) + f\right] + \sqrt{2D} \, \xi \ .
\end{equation}
Here, $\gamma$ is the friction coefficient of the cargo (with dimensions of mass/time), which experiences an external load force $f$, a force from the coupling to the motor (at position $y$) with stiffness $\kappa$, and a random force due to thermal fluctuations with strength quantified by diffusion coefficient $D = k_{\rm B}T/\gamma$. The uncorrelated Gaussian noise $\xi(t)$ has zero mean and unit variance. At steady state, the heat flow into the cargo is~\cite{leighton2023inferring}
\begin{equation}
\label{eq:heatdefinition}
    \dot{Q}_X \equiv \left\langle - \left[\kappa(y-x) + f\right] \circ \dot{x} \right\rangle ,
\end{equation}
where the angle brackets denote an ensemble average with respect to the steady-state distribution, and ``$\circ$" denotes multiplication in the Stratonovich sense~\cite{seifert2012stochastic}. At steady state, the temporal evolution is independent of initial conditions; in particular, the distribution of the relative position $x - y$ does not depend on time.

Directly calculating the heat flow $\dot{Q}_X$ requires simultaneous measurements of both the motor and cargo positions along trajectories. Although each of these measurements can be performed independently~\cite{schnitzer1997kinesin,wirth2023minflux}, no experiment has yet measured them simultaneously. Thus, we develop an alternate method for estimating the heat flow from experimentally accessible measurements.

We show that the heat flow $\dot{Q}_X$ can be related to the statistics of cargo displacements $\Delta x$ under only mild assumptions on the dynamics of the motor $Y$. We start by explicitly integrating Eq.~\eqref{eq:xdynamics} to get the cargo displacement $\Delta x \equiv x(t + \Delta t) - x(t)$, given $x(t) = x$ and $y(t) = y$, for a timestep $\Delta t$ sufficiently small that the motor is roughly stationary, giving the distribution
\begin{align}
\label{eq:deltaxdistribution}
p(&\Delta x \, | \, x,y)\\
& = \mathcal{N}\Big[\left(y\!-\!x+f/\kappa\right)\!\left(1\!-\!e^{-\Delta t/\tauR} \right), \sigma^2\!\left( 1\!-\!e^{-2\Delta t\tauR} \right)\! \Big]\,, \nonumber
\end{align}
for cargo relaxation time $\tauR = \gamma/\kappa$ and variance $\sigma^2 = D \tauR = k_\mathrm{B}T/\kappa$, given a stationary motor $Y$.

From (\ref{eq:deltaxdistribution}), the cargo's mean squared displacement (MSD) $\langle \Delta x^2 \rangle$ over timestep $\Delta t$ can be obtained by first averaging over $p(\Delta x\,|\,x,y)$, then averaging over the appropriate NESS describing the joint $X$-$Y$ system. Even though we generally do not know this steady-state distribution, we can substitute for the cargo heat flow (\ref{eq:heatdefinition}):
\begin{equation}\label{eq:MSD}
\begin{aligned}
\!\langle \Delta x^2 \rangle_{\mathrm{neq}} = \sigma^2\!\left[2\left(1\!-\!e^{-\Delta t/\tauR}\right) \! - \! \beta\dot Q_X\tauR \left(1\!-\!e^{-\Delta t/\tauR} \right)^2 \right]. 
\end{aligned}
\end{equation}
The LHS can be estimated through an empirical average of squared displacements $\Delta x^2$ measured over many time intervals of duration $\Delta t$. In principle, then, (\ref{eq:MSD}) provides a means to infer the heat flow $\dot{Q}_X$ and thus the mechanism of the motor $Y$, using only the experimentally accessible cargo MSD. However, directly applying (\ref{eq:MSD}) requires the parameters $\tauR$ and $\sigma$, which depend on the cargo diffusion coefficient and linker stiffness.

This last requirement can be removed by considering also the MSD of the cargo in an equilibrium state, where the motor does not undergo directed motion (this could be achieved, for example, by immobilizing the motor or eliminating the chemical-potential difference of its fuel). At equilibrium, the cargo exchanges no net energy with the thermal bath ($\dot{Q}_X = 0$), with corresponding MSD $\langle \Delta x^2 \rangle_{\mathrm{eq}} = 2\sigma^2 (1 - e^{-\Delta t/\tauR})$ obtained from (\ref{eq:MSD}). Dividing the nonequilibrium and equilibrium MSDs and Taylor expanding to first order in the timestep $\Delta t$ yields
\begin{equation} \label{eq:msdratiotaylor}
    \frac{\langle \Delta x^2 \rangle_{\mathrm{neq}}}{\langle\Delta x^2 \rangle_{\mathrm{eq}}} = 1 - \tfrac{1}{2} \beta \dot{Q}_X \Delta t + \mathcal{O}(\Delta t^2)\ .
\end{equation}
This result is independent of all parameters characterizing the cargo-motor linker and cargo diffusivity. The cargo heat flow can thus be inferred using the estimator
\begin{equation} \label{eq:heatestimator}  
    \beta\widehat{\dot{Q}_X} \equiv \frac{2}{\Delta t} \left( 1- \frac{\langle \Delta x^2 \rangle_{\mathrm{neq}}}{\langle\Delta x^2 \rangle_{\mathrm{eq}}} \right)\ ,
\end{equation}
which is equal to the heat flow in the limit of small $\Delta t$. Equation~\eqref{eq:heatestimator} is our main result, relating the cargo heat flow to experimentally accessible quantities that can be determined by observing only the cargo. Thus, we can infer whether the motor acts as a Maxwell demon without measuring the motor dynamics. Further, the sign of the cargo heat flow (and hence the mode of operation of the motor) can be determined by simply comparing the nonequilibrium cargo MSD to the equilibrium cargo MSD: Maxwell-demon behavior corresponds to a cargo MSD that is smaller than the equilibrium MSD.

The bias and precision of the heat estimator~\eqref{eq:heatestimator} determine its practical utility. The bias can be computed by expanding \eqref{eq:heatestimator} in powers of $\Delta t$ and comparing to the true heat (SI~\ref{SI:bias}). The estimator is unbiased as $\Delta t\to0$, with a first-order correction $-\beta\dot{Q}_X \,\Delta t /(2\tauR)$ proportional to the true heat flow. The relative bias is thus $-\Delta t/(2\tauR)$; accurate estimation requires $\Delta t \ll \tauR$.

The estimator's variance depends on the fourth moments of the equilibrium and nonequilibrium $\Delta x$ distributions (SI~\ref{SI:variance}). When the marginal distribution of $x\!-\!y$ is Gaussian, the result simplifies to $16/\left(N\Delta t^2\right) + \mathcal{O}(\Delta t^{-1})$, for $N$ trajectory increments of duration $\Delta t$. 

These results are exact in the limit of small $\Delta t$, and in SI~\ref{SI:exactcalcs} we verify them analytically in simple models of conventional and information engines. In simulations of more complex systems, there can be additional contributions to bias and variance.

\emph{Extensions}---Here we briefly describe extensions of the main result (\ref{eq:heatestimator}). The heat estimator is shown to be remarkably robust in situations where a measurement of the cargo position contains non-negligible noise satisfying reasonable conditions (SI~\ref{sec:SImesnoise}). In particular, the sign of the heat flow (and hence the motor's operational mode) can be inferred with or without knowing the variance $\sigma_{\mathrm{m}}^2$ of the measurement noise, while the exact heat flow can still be inferred when $\sigma_{\mathrm{m}}^2$ is known.

In SI~\ref{SI:noneqnoise} we further show how the heat estimator can be extended to certain classes of nonequilibrium noise, and thus to systems with active fluctuations. Similar to the case with measurement noise, the relative magnitudes of the equilibrium and nonequilibrium MSDs allow inference of the heat flow's direction even when the nonequilibrium noise strength is not precisely known.

\emph{Colloidal particle under feedback control}---
A simple setup for benchmarking our estimator is the colloidal engine~\cite{saha2021maximizing, Lucero2021_Maximal, Saha2022_Bayesian, Saha2022_Information} reviewed in~\cite{du2024performance} and illustrated in Fig.~\ref{fig:fig1}a and b. Here, a micron-scale bead ($X$) is manipulated via optical tweezers (with trap center $Y$) that exert a linear vertical force on the bead, parallel to gravity. We simulate this setup, where the trap center is updated according to a feedback rule, raising the bead against gravity and storing gravitational free energy. 

The engine can operate as a Maxwell demon (Fig.~\ref{fig:fig1}b), where favorable thermal fluctuations are rectified by moving the trap only when the bead fluctuates above the trap center, without doing explicit work on the bead. In this scenario, heat from the thermal bath is rectified to impart an average upward velocity to the bead, with $\dot{Q}_X > 0$. Alternatively, in the conventional-engine mode (Fig.~\ref{fig:fig1}a), gravitational free energy is stored by raising the trap center at a rate independent of the bead position. In this case, work is done on the bead, with the bead dissipating heat ($\dot{Q}_X < 0$) to its environment via frictional drag with the surrounding fluid. 

The bead dynamics obey (\ref{eq:xdynamics}) with constant force $f=-m g$ and bead effective mass $m$ (accounting for buoyancy). The bead position $x_n$ is recorded at intervals of 20~\textmu s, and the trap position is updated in accordance with a feedback rule at the next timestep, hence with feedback delay 20~\textmu s. In the Maxwell-demon operation mode, the trap center is updated according to
\begin{equation}
    y_{n+1} = y_n + \Theta(x_n - y_n)\,  \alpha (x_n - y_n)\ ,
\end{equation}
with $\Theta$ the Heaviside step function and the (constant) feedback gain $\alpha$ chosen so that the trap does, on average, zero work on the bead~\cite{saha2021maximizing,Saha2022_Bayesian}. In the conventional-engine mode, the trap center is simply shifted a constant distance every 20~\textmu s, sufficiently rapidly that the bead position is effectively constant during the trap update. 

\begin{figure}[t]
\centering
\includegraphics[width=0.8\linewidth]{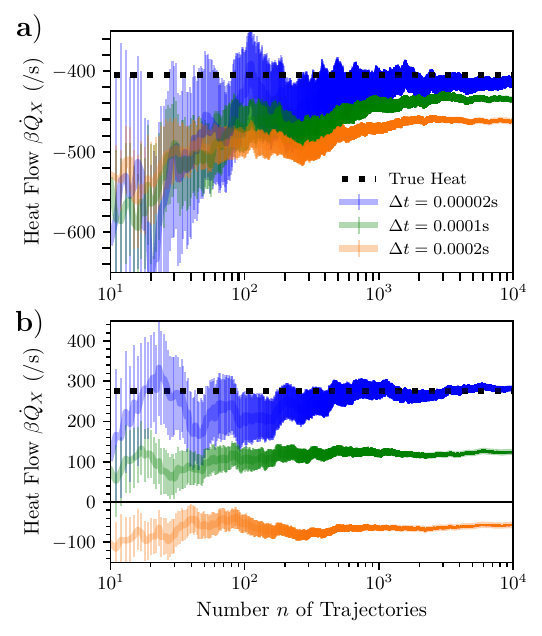}
\caption{The heat estimator~\eqref{eq:heatestimator} applied to the colloidal-engine simulation data, showing mean and standard error (error bars) as a function of the number $n$ of trajectories of duration $1$\,s, for different sampling times $\Delta t$. Operation modes are a) conventional engine and b) Maxwell demon. Black dotted lines: true heat flows (calculated directly from simulations). Simulation parameters are $\gamma \approx 2.6 \times 10^{-8}\, \mathrm{N\cdot s/m}$, $\kappa \approx 3.52 \times 10^{-5} \, \mathrm{N/m}$, $m = 1.4137 \times 10^{-14}\, \mathrm{kg}$.}
\label{fig:fig2}
\end{figure}

Figure~\ref{fig:fig2} illustrates the estimator's performance, confirming the behavior of the bias and variance in SI~\ref{SI:bias} and \ref{SI:variance}: the estimator converges in the large-data limit to a value determined by the timestep $\Delta t$, with $\widehat{\dot{Q}_X} \rightarrow \dot{Q}_X$ as $\Delta t \rightarrow 0$. For a given quantity of data, the estimator has larger variance for smaller $\Delta t$ (when more details of the $X$ dynamics are resolved), in accordance with \eqref{eq:estimatorvariance}. Conversely, increasing $\Delta t$ averages out more of the $X$ dynamics, decreasing the variance and increasing the bias.

For the smallest sampling time ($\Delta t\!=\!20$~\textmu s) and $10^4$ trajectories of duration $1$\,s, the estimator comes within $5$\% of the true heat flow for both Maxwell-demon ($<$$0.5$\%) and conventional-engine ($\approx$$3$\%) modes. This matches theoretical predictions (\ref{eq:estimatorbias}) of the magnitude of the estimator's bias ($\Delta t /2\tauR \approx 0.014$ for the parameters used) and demonstrates that the heat flow can be practically estimated, given experimentally feasible quantities of data and sampling time $\Delta t$.

\emph{Kinesin pulling diffusive cargo}---We now demonstrate the heat estimator's effectiveness in molecular motors. As a paradigmatic example, let a kinesin motor ($Y$) pull a diffusive cargo ($X$) against an external force $f$. We explicitly model in SI~\ref{SI:kinesinmodel} the kinesin motor dynamics using a two-state discrete model~\cite{ariga2018nonequilibrium}. We combine these motor dynamics with continuous cargo dynamics~\eqref{eq:xdynamics} to obtain bipartite dynamics. We numerically simulate this model with experimentally determined~\cite{ariga2018nonequilibrium} parameter values to obtain sample trajectories with experimentally accessible spatiotemporal resolution, numbers of trajectories, and $1$\,s duration. We show in SI~\ref{SI:morekinesinresults} that the estimator can precisely and accurately infer heat flow from kinesin simulations. For $\Delta t\approx 50$~\textmu s, $\mathcal{O}\left(10^3\right)$ trajectories give precise estimates of the heat flow.
Such numbers are approached in recent single-molecule experiments~\cite{peters20263D-MINSTED}.

\begin{figure}[h]
    \centering
    \includegraphics[width =0.8\linewidth]{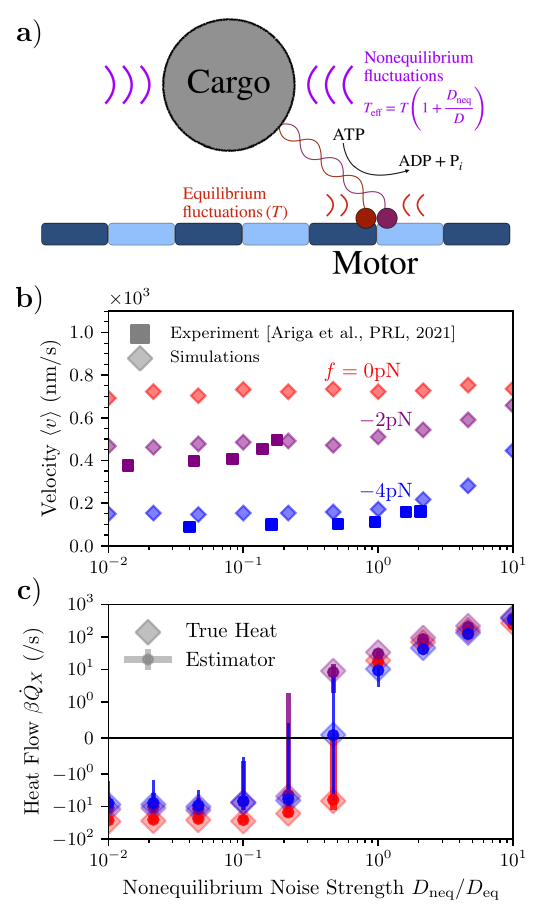}
    \caption{Inferring heat flows for a kinesin motor pulling a diffusive cargo. a) Schematic illustrating the nonequilibrium noise applied to the cargo. b) Transport velocity $\langle v\rangle$ as a function of nonequilibrium noise strength $D_\mathrm{neq}/D_\mathrm{eq}$ for different external forces $f=0$pN (red), $f=-2$pN (purple), and $f=-4$pN (blue). Diamonds: simulations; filled squares: experimental data from Ref.~\cite{ariga2021noise}. Uncertainties are smaller than the points. c) Heat flow as a function of nonequilibrium noise strength $D_\mathrm{neq}/D_\mathrm{eq}$ for different external forces. Diamonds indicate the true heat flow, while points with error bars indicate the mean and SEM of the heat estimator with $\Delta t = 5\times 10^{-5}$\,s and $n=4,000$. We use $D_\mathrm{eq} = 126,680$\, nm$^2$/s. The y-axis in c) is a symmetric log axis, with a linear scale between $-1$ and 1 and a logarithmic scale otherwise.}
    \label{fig:fig3}
\end{figure}

For a wide range of physiologically plausible parameter values and for equilibrium thermal fluctuations acting on the cargo, we find that the cargo heat flow is negative ($\dot{Q}_X<0$), i.e., kinesin operates as a conventional engine. Motivated by the experimental finding of faster kinesin operation under applied nonequilibrium noise~\cite{ariga2021noise} and theoretical results showing information flows are required to take advantage of different sources of fluctuations~\cite{leighton2024information}, we modify the cargo dynamics to include nonequilibrium noise, adding a term $\sqrt{2D_\mathrm{neq}}\,\xi_\mathrm{neq}(t)$ to Eq.~\eqref{eq:xdynamics}, as illustrated in Fig.~\ref{fig:fig3}a. Here, $\xi_\mathrm{neq}(t)$ is Gaussian white noise with mean 0 and variance 1, and $D_\mathrm{neq}$ quantifies the nonequilibrium noise strength. The added nonequilibrium noise is mathematically equivalent to increasing the effective temperature of the cargo from $T$ to $T_\mathrm{eff} = T\left(1 + D_\mathrm{neq}/D\right)$. Such an effective temperature can also be derived following Ref.~\cite{sorkin2024second}.

Figure~\ref{fig:fig3}b shows the velocity as a function of the nonequilibrium noise strength $D_\mathrm{neq}$; consistent with the experimental findings of Ref.~\cite{ariga2021noise}, in our simulations velocity increases with $D_\mathrm{neq}$, with larger increases for larger external forces opposing the motor motion ($f<0$). Figure~\ref{fig:fig3}c shows that the corresponding cargo heat flow $\dot{Q}_X$ is negative for small $D_\mathrm{neq}\ll D_\mathrm{eq}$ but crosses over to positive values slightly before $D_\mathrm{neq}=D_\mathrm{eq}$. For $D_\mathrm{neq}>D_\mathrm{eq}$, this kinesin model operates as a Maxwell demon, with $\dot{Q}_X>0$. For experimentally accessible quantities of simulation data, the heat estimator accurately and precisely captures the transition to Maxwell-demon behavior.

\emph{Discussion}---In this Letter, we introduced a new statistical estimator~\eqref{eq:heatestimator} for subsystem heat flows in multicomponent systems, using trajectory measurements of only a single degree of freedom. The estimator works remarkably well, precisely and accurately estimating heat flows from experimentally accessible quantities of data for a simulated colloidal information engine that was previously experimentally realized and a simulated kinesin motor. Thus we can reliably detect Maxwell-demon behavior in real systems. We find kinesin pulls cargo faster in the presence of nonequilibrium noise, with the heat flow strikingly changing sign (indicating Maxwell-demon behavior) in the regime where the velocity increases. Our simulations show this should be experimentally accessible with $\approx\!4,\!000$ trajectories of duration $1$~s.

Our heat-estimator derivation assumed that the cargo evolves according to linear Langevin dynamics. While this assumption is valid for the systems considered here, generalizing to nonlinear forces is an interesting future direction. Similarly, the nonequilibrium white noise considered is the simplest model for active fluctuations in the cell; it would be interesting to explore temporally correlated fluctuations. While we considered only a single kinesin motor, \textit{in vivo} transport often involves multiple motors. Building on the stochastic thermodynamics of collective motor-driven transport~\cite{leighton2022performance,leighton2022dynamic}, the estimator should be straightforwardly applicable to estimate cargo heat flows in multi-motor systems.

Our results suggest that molecular motors such as kinesin behave differently in a nonequilibrium environment, in this case strongly outperforming their behavior in equilibrium environments by using qualitatively different modes of thermodynamic operation. Perhaps the molecular machinery of the cell has evolved to take advantage of the nonequilibrium fluctuations inherent in its environment. Exploring this hypothesis and its consequences will be of great interest.

\emph{Acknowledgments}---We thank Takayuki Ariga (Osaka University) for helpful discussions, and Henry Mattingly (Flatiron Institute) and Bezia Lemma and Linnea Lemma (Princeton University) for feedback on the manuscript. This work was supported by an Alexander von Humboldt Foundation Research fellowship (J.d.B); Natural Sciences and Engineering Research Council of Canada (NSERC) doctoral and postdoctoral fellowships (M.P.L.); a Yale Mossman postdoctoral fellowship (M.P.L.); grant FQXi-IAF19-02 from the Foundational Questions Institute Fund, a donor-advised fund of the Silicon Valley Community Foundation (J.B.\ and D.A.S.); an NSERC Discovery Grant and Discovery Accelerator Supplement RGPIN-2020-04950 (D.A.S.); and a Tier-II Canada Research Chair CRC-2020-00098 (D.A.S.).

\emph{Code and Data Availability}---Our simulation code for both the colloidal particle under feedback control, and the kinesin molecular motor, is available on Github~\cite{github}.

\bibliography{main}

\end{document}


\title{Supplementary Material for ``Hunting for Maxwell's Demon in the Wild"}

\author{Johan du Buisson}
\thanks{These authors contributed equally.}
\affiliation{Department of Physics, Simon Fraser University, Burnaby, BC, V5A 1S6, Canada.}

\author{Jannik Ehrich}
\thanks{These authors contributed equally.}
\affiliation{Department of Physics, Simon Fraser University, Burnaby, BC, V5A 1S6, Canada.}

\author{Matthew P.\ Leighton}
\thanks{These authors contributed equally.}
\affiliation{Department of Physics, Simon Fraser University, Burnaby, BC, V5A 1S6, Canada.}
\affiliation{Department of Physics and Quantitative Biology
Institute, Yale University, New Haven, CT, 06511, USA}

\author{Avijit Kundu}
\affiliation{Department of Physics, Simon Fraser University, Burnaby, BC, V5A 1S6, Canada.}

\author{Tushar K.\ Saha}
\affiliation{Department of Physics, Simon Fraser University, Burnaby, BC, V5A 1S6, Canada.}
\affiliation{Current Address: MKS Instruments, Inc., 130-13500 Verdun Place, Richmond, BC, V6V 1V2, Canada}

\author{John Bechhoefer}
\affiliation{Department of Physics, Simon Fraser University, Burnaby, BC, V5A 1S6, Canada.}

\author{David A.\ Sivak}%
\email{dsivak@sfu.ca}
\affiliation{Department of Physics, Simon Fraser University, Burnaby, BC, V5A 1S6, Canada.}

\maketitle

\setcounter{secnumdepth}{4}

\section{Additional details and extensions of the heat estimator}

\subsection{Detailed derivation of the heat estimator}
\label{SI:derivation}
Here we show the calculations leading to the result (\ref{eq:MSD}), relating the experimentally measurable MSD of the bead to the steady-state heat flow without requiring knowledge of the steady-state distribution $p(x,y)$. First, we rewrite the definition of heat (\ref{eq:heatdefinition}) in a more convenient form. Given a 
(generally unknown) 
steady-state distribution $p(x,y)$ and the Stratonovich convention,
\begin{subequations}
\begin{align} 
    \dot{Q}_X &= \langle -[\kappa(y - x) + f] \circ \dot{x} \rangle \\
    & = \int\hspace{-0.5em} \int \mathrm{d}x \, \mathrm{d}y \, \left[\kappa(x - y) - f \right]J_X(x,y) \label{eq:heatStrato} \ .
\end{align}
\end{subequations}
Substituting the probability current's $X$-component
\begin{equation}
    J_X(x,y) = \frac{\kappa(y - x) + f}{\gamma}p(x,y) - D\, \partial_x p(x,y)
\end{equation}
gives
\begin{equation}
    \dot{Q}_X = -\frac{\langle [\kappa(y - x) + f]^2\rangle_{p(x,y)}}{\gamma} - \int \dd{x} \dd{y} \, \left[\kappa(x - y) - f \right] D \, \partial_x p(x,y) \ .
\end{equation}
Using integration by parts and the normalization of $p(x,y)$, this reduces to 
\begin{equation} \label{eq:heatredef}
\dot{Q}_X = -\frac{\langle [\kappa(y - x) + f]^2\rangle_{p(x,y)}}{\gamma} + \kappa D \ .
\end{equation}

To relate the bead's MSD for timestep $\Delta t$ to the heat flow, we consider a timestep sufficiently short that the motor ($Y$) can be regarded as static.  In that limit, 
\begin{equation}
    \langle \Delta x^2\rangle_{\textnormal{neq}} \equiv \langle \Delta x^2 \rangle_{p(\Delta x, x, y)} = \left \langle \langle \Delta x^2 \rangle_{p(\Delta x\,|\,x,y)} \right\rangle_{p(x,y)}\ .
\end{equation}
Using (\ref{eq:deltaxdistribution}) to substitute for $p(\Delta x\,|\,x,y)$ gives
\begin{equation}
    \langle \Delta x^2\rangle_{p(\Delta x\,|\,x,y)} = \left[y - x + f/\kappa \right]^2 \left(1 - e^{-\Delta t/\tauR} \right)^2 + \sigma^2 \left(1 - e^{-2\Delta t\tauR} \right)\ .
\end{equation}
Averaging with respect to $p(x,y)$ gives
\begin{equation}
    \langle \Delta x^2 \rangle_{\mathrm{neq}} = \left\langle \left[y - x + f/\kappa \right]^2 \right \rangle_{p(x,y)}\left(1 - e^{-\Delta t/\tauR} \right)^2 + \sigma^2 \left(1 - e^{-2\Delta t/\tauR} \right)\ ,
\end{equation}
or, in terms of the steady-state heat flow (\ref{eq:heatredef}), 
\begin{equation}\label{eq:MSDneq}
    \langle \Delta x^2 \rangle_{\mathrm{neq}} = \frac{\tauR}{\kappa} \left(-\dot{Q}_X + \kappa D \right)\left(1 - e^{-\Delta t/\tauR} \right)^2 + \sigma^2 \left(1 - e^{-2\Delta t/\tauR} \right)\ .
\end{equation}
Substituting $D\tauR = \sigma^2 = k_{\mathrm{B}}T/\kappa$ yields the result \eqref{eq:MSD}:
\begin{equation} \label{eq:MSDneq2}
    \langle \Delta x^2 \rangle_{\mathrm{neq}} = \sigma^2\left[2\left(1-e^{-\Delta t/\tauR}\right) - \beta\dot Q_X \tauR \left(1\!-\!e^{-\Delta t/\tauR} \right)^2 \right]\ .
\end{equation}

\subsection{Bias of the heat estimator}
\label{SI:bias}
As mentioned in the main-text section \emph{Subsystem heat-flow estimator}, 
the heat estimator~\eqref{eq:heatestimator} is unbiased up to first order in $\Delta t$. To show this, we insert into the estimator \eqref{eq:heatestimator} the full expressions for the MSDs \eqref{eq:MSDneq} and $\langle \Delta x^2 \rangle_{\mathrm{eq}} = 2\sigma^2 (1 - e^{-\Delta t/\tauR})$, expand around $\Delta t=0$, and subtract the true heat flow from both sides to obtain
\begin{equation}\label{eq:estimatorbias}
\beta\widehat{\dot{Q}_X} - \beta\dot{Q}_X = -\frac{1}{2\tauR}
\beta\dot{Q}_X \Delta t + \mathcal{O}(\Delta t^2).
\end{equation}

\subsection{Precision of the heat estimator}
\label{SI:variance}
The precision of the heat estimator requires a more involved calculation. We begin by considering the estimator for the MSD,
\begin{equation}
\widehat{\left\langle \Delta x^2\right\rangle} = \frac{1}{N}\sum_{i=1}^N\Delta x_i^2\ .
\end{equation}
The variance of this estimator is
\begin{subequations}
\begin{align}
\mathrm{Var}\left(\widehat{\left\langle \Delta x^2\right\rangle}\right) & = \left\langle \left( \widehat{\left\langle \Delta x^2\right\rangle} - \left\langle \widehat{\left\langle \Delta x^2\right\rangle}\right\rangle\right)^2\right\rangle\\
& = \left\langle \left( \left[\frac{1}{N}\sum_{i=1}^N\Delta x_i^2\right] - \left\langle \Delta x^2\right\rangle\right)^2\right\rangle\\
& = \left\langle \left(\frac{1}{N}\sum_{i=1}^N\Delta x_i^2\right)^2\right\rangle - 2\left\langle \frac{1}{N}\sum_{i=1}^N\Delta x_i^2 \left\langle\Delta x^2\right\rangle\right\rangle + \left\langle \left\langle\Delta x^2\right\rangle^2\right\rangle \label{eq:trickyLine} \\
& = \frac{1}{N^2}\left[ N\left\langle \Delta x^4\right\rangle + (N^2-N)\left\langle \Delta x^2\right\rangle^2\right] - 2\left\langle \Delta x^2\right\rangle^2 + \left\langle \Delta x^2\right\rangle^2\\
& = \frac{1}{N}\left( \left\langle \Delta x^4\right\rangle -\left\langle \Delta x^2\right\rangle^2 \right)\ .
\end{align}
\end{subequations}
In \eqref{eq:trickyLine} we assumed that if $i\neq j$, steps $\Delta x_i$ and $\Delta x_j$ are independent and identically distributed random variables, in order to expand the squared sum into terms of the form $\langle \Delta x_i^4\rangle$ and $\langle \Delta x_i^2\rangle\langle \Delta x_j^2\rangle = \langle \Delta x^2\rangle^2$. 

Contributions from both the equilibrium and nonequilibrium MSD's must be considered. The standard deviation of the heat estimator is
\begin{subequations}
\begin{align}
\mathrm{Std}\left(\beta\widehat{\dot{Q}_X}\right) & = 
\left[\left( \frac{\partial \beta\widehat{\dot{Q}_X}}{\partial \left\langle \Delta x^2\right\rangle_\mathrm{eq}}\right)^2 \mathrm{Var}\left(\widehat{\left\langle \Delta x^2\right\rangle}_\mathrm{eq}\right) + \left( \frac{\partial \beta\widehat{\dot{Q}_X}}{\partial \left\langle \Delta x^2\right\rangle_\mathrm{neq}}\right)^2 \mathrm{Var}\left(\widehat{\left\langle \Delta x^2\right\rangle}_\mathrm{neq}\right)\right]^{1/2}\\
& = \left[\left(\frac{2}{\Delta t}\cdot\frac{\left\langle \Delta x^2\right\rangle_\mathrm{neq}}{\left\langle \Delta x^2\right\rangle_\mathrm{eq}^2}\right)^2 \left(\frac{1}{N}\left\langle \Delta x^4\right\rangle_\mathrm{eq}-\frac{1}{N}\left\langle\Delta x^2\right\rangle^2_\mathrm{eq}\right) + \left(\frac{-2}{\Delta t\cdot\left\langle \Delta x^2\right\rangle_\mathrm{eq}}\right)^2\left(\frac{1}{N}\left\langle \Delta x^4\right\rangle_\mathrm{neq}-\frac{1}{N}\left\langle\Delta x^2\right\rangle^2_\mathrm{neq}\right) \right]^{1/2}\\ 
& = \frac{2}{\sqrt{N} \Delta t}\cdot\frac{1}{\left\langle \Delta x^2\right\rangle_\mathrm{eq}} \left[ \left(\frac{\left\langle \Delta x^2\right\rangle_\mathrm{neq}}{\left\langle \Delta x^2\right\rangle_\mathrm{eq}}\right)^2\left(\left\langle \Delta x^4\right\rangle_\mathrm{eq}-\left\langle\Delta x^2\right\rangle^2_\mathrm{eq}\right) + \left(\left\langle \Delta x^4\right\rangle_\mathrm{neq}-\left\langle\Delta x^2\right\rangle^2_\mathrm{neq}\right)\right]^{1/2}\ .
\end{align}
\end{subequations}
If $\Delta x$ is Gaussian-distributed (once marginalized over $p(x,y)$), then this simplifies significantly to
\begin{equation}\label{eq:estimatorvariance}
\mathrm{Std}\left(\beta\widehat{\dot{Q}_X}\right) = \frac{4}{\sqrt{N}\Delta t} + \mathcal{O}\left(\Delta t^{-1/2}\right)\ .
\end{equation}

\subsection{Heat estimator with measurement noise}
\label{sec:SImesnoise}
Here we mathematically justify the statements in the main-text section \emph{Extensions} regarding extensions of the heat estimator. We consider the case where the error in measurements of the bead position is not negligible, with the $n$th measurement $z_n$ related to the true bead position $x_n$ by
\begin{equation} \label{eq:measurementnoise}
    z_n = x_n + \nu_n\ ,
\end{equation}
with the $\nu_n$ being independent and identically distributed random variables with mean $0$ and variance $\sigma_{\mathrm{m}}$. Using (\ref{eq:measurementnoise}), the estimator (\ref{eq:heatestimator}) can be written as
\label{SI:measurement}
\begin{equation} \label{eq:heatestimator2}  
\beta\widehat{\dot Q_X} = \frac{2}{\Delta t} \left( 1- \frac{{\langle\Delta z^2 \rangle}_{\mathrm{neq}} - 2\sigma_{\mathrm{m}}^2}{{\langle\Delta z^2 \rangle}_{\mathrm{eq}}- 2\sigma_{\mathrm{m}}^2} \right)\ ,
\end{equation}
meaning that the true heat can still be estimated by comparing the measured bead MSD $\langle \Delta z^2\rangle$ in nonequilibrium and equilibrium, provided that the variance $\sigma_{\mathrm{m}}^2$ of the measurement noise is known and assuming that the nature of the measurement noise is the same under both nonequilibrium and equilibrium conditions.

When the variance of the measurement noise is not known, we can nevertheless still bound the heat flows by comparing the observed bead MSD in nonequilibrium and equilibrium. To show this, we Taylor expand (\ref{eq:heatestimator2}) with respect to $\sigma_{\mathrm{m}}$, obtaining
\begin{equation}
\beta\widehat{\dot Q_X} = \frac{2}{\Delta t} \left[1 - \frac{\langle\Delta z^2 \rangle_{\mathrm{neq}}}{ \langle\Delta z^2 \rangle_{\mathrm{eq}}}  - \left({\langle\Delta z^2 \rangle}_{\mathrm{neq}} - {\langle\Delta z^2 \rangle}_{\mathrm{eq}}\right) \sum_{n = 1}^{\infty} \frac{2^n \sigma_{\mathrm{m}}^{2n}}{{\langle\Delta z^2 \rangle}_{\mathrm{eq}}^{n + 1}}\right]\ .
\end{equation}
Since every term inside the sum in the above expression is necessarily positive, the sign of the term
\begin{equation}
\left({\langle\Delta z^2 \rangle}_{\mathrm{neq}} -{\langle\Delta z^2 \rangle}_{\mathrm{eq}}\right) \sum_{n = 1}^{\infty} \frac{2^n \sigma_{\mathrm{m}}^{2n}}{{\langle\Delta z^2 \rangle}_{\mathrm{eq}}^{n + 1}}
\end{equation}
depends only on the sign of $\langle\Delta z^2 \rangle_{\mathrm{neq}} -{\langle\Delta z^2 \rangle}_{\mathrm{eq}}$. In particular, if the measured nonequilibrium MSD exceeds that in equilibrium, then
\begin{equation}
    \beta \widehat{\dot Q_X} < \frac{2}{\Delta t} \left( 1- \frac{{\langle\Delta z^2 \rangle}_{\mathrm{neq}}}{{\langle\Delta z^2 \rangle}_{\mathrm{eq}}} \right) < 0\ .
    \label{eq:bound1}
\end{equation}
Otherwise,
\begin{equation}
\beta\widehat{\dot Q_X} > \frac{2}{\Delta t} \left( 1- \frac{{\langle\Delta z^2 \rangle}_{\mathrm{neq}}}{{\langle\Delta z^2 \rangle}_{\mathrm{eq}}} \right) > 0\ .
\label{eq:bound2}
\end{equation}
Thus, the heat flow is bounded in terms of only the observable (noisy) equilibrium and nonequilibrium MSDs. Additionally, these two bounds~\eqref{eq:bound1} and \eqref{eq:bound2} imply that if the measured cargo MSD is larger under nonequilibrium conditions than in equilibrium, then the subsystem $Y$ behaves as a conventional engine; otherwise, it behaves as a Maxwell demon.

\subsection{Direct heat estimator and its properties}
\label{SI:directestimator}
A key feature of heat estimator~\eqref{eq:heatestimator} is that it does not require knowledge of the parameters that govern the dynamics of the cargo $X$, relying instead on comparing measured MSDs for nonequilibrium and equilibrium dynamics. If, however, these parameters (specifically the cargo diffusivity $D_\mathrm{eq}$ and linker stiffness $\kappa$) are known, then \eqref{eq:MSDneq2} can be rearranged to derive a \emph{direct heat estimator} that requires only the nonequilibrium MSD:
\begin{equation}\label{eq:directestimator}
\beta\widehat{\dot{Q}_X}^\mathrm{(dir)} = \frac{1}{\tauR} + \frac{1}{\tauR} \left(1-e^{-\Delta t/\tauR}\right)^{-2}\left[\left(1-e^{-2\Delta t/\tauR}\right)-\frac{\left\langle \Delta x^2\right\rangle_\mathrm{neq}}{\sigma^2}\right]\ .
\end{equation}
This alternate estimator is unbiased for all time intervals $\Delta t$ sufficiently short that the motor is roughly stationary.

\subsection{Heat estimator with nonequilibrium noise}
\label{SI:noneqnoise}
In the main-text 
section \emph{Kinesin pulling diffusive cargo}, 
we consider kinesin pulling a cargo with additional nonequilibrium noise, such that the cargo dynamics follow the overdamped Langevin equation
\begin{equation}
\dot{x} = \gamma^{-1}\left[f + \kappa (y-x)\right] + \underbrace{\sqrt{2D_\mathrm{eq}}\,\xi(t)}_{\mathrm{equilibrium}\,\mathrm{noise}} + \underbrace{\sqrt{2D_\mathrm{neq}}\,\xi_\mathrm{neq}(t)}_{\mathrm{nonequilibrium}\,\mathrm{noise}}\ .
\end{equation}
The nonequilibrium noise is characterized by a magnitude $D_\mathrm{neq}$, and $\xi_\mathrm{neq}(t)$ is Gaussian white noise with zero mean and unit variance. We combine the effects of the equilibrium and nonequilibrium noises into a single term,
\begin{equation}
\sqrt{2D_\mathrm{eq}}\,\xi(t) + \sqrt{2D_\mathrm{neq}}\,\xi_\mathrm{neq}(t) = \sqrt{2D_\mathrm{eff}}\,\xi_\mathrm{tot}(t)\ ,
\end{equation}
for effective diffusion coefficient $D_\mathrm{eff}\equiv D_\mathrm{eq} + D_\mathrm{neq}$. Using the fluctuation-dissipation relation $\gamma D_\mathrm{eq} = k_\mathrm{B} T$, we define an effective temperature 
\begin{subequations}
\begin{align}
T_\mathrm{eff} & \equiv \frac{1}{k_\mathrm{B}}\gamma D_\mathrm{eff}\\
& = T\left(1 + \frac{D_\mathrm{neq}}{D_\mathrm{eq}}\right)\ .
\end{align}
\end{subequations}
Applying the heat estimator gives
\begin{equation}
\frac{1}{k_\mathrm{B}T_\mathrm{eff}}\widehat{\dot{Q}_X} = \frac{2}{\Delta t} \left( 1- \frac{\langle \Delta x^2 \rangle_{\mathrm{neq}}}{\langle\Delta x^2 \rangle_{\mathrm{eq}}} \right)\ ,
\end{equation}
so that the heat scaled by the true temperature is
\begin{equation}
\beta \widehat{\dot{Q}_X} = 
\frac{T_{\rm eff}}{T}
\frac{2}{\Delta t} \left( 1- \frac{\langle \Delta x^2 \rangle_{\mathrm{neq}}}{\langle\Delta x^2 \rangle_{\mathrm{eq}}} \right)\ .
\end{equation}
Here $\langle \Delta x^2\rangle_\mathrm{neq}$ and $\langle \Delta x^2\rangle_\mathrm{eq}$ are the mean squared displacements when $Y$ respectively does and does not impart nonequilibrium driving forces to $X$. The nonequilibrium noise on $X$ is present in both cases. When nonequilibrium noise is present but the precise value of $T_\mathrm{eff}$ is not known, comparing the equilibrium and nonequilibrium MSDs still permits correct inference of the sign of $\dot{Q}_X$, and thus the direction of the heat flow.

\section{Exact calculations for simple systems}
\label{SI:exactcalcs}
Here we detail simple models where the heat flow and estimator can be calculated exactly, as mentioned at the end of main-text section \emph{Subsystem heat-flow estimator}.

\subsection{Constant-velocity conventional engine}
\label{SI:conventionalexact}
Consider a constant-velocity conventional engine pulling a diffusive cargo, with respective positions $y(t)$ and $x(t)$. The dynamics are given by
\begin{subequations}
\begin{align}
\dot{x} & = \frac{\kappa}{\gamma}(y - x) + \sqrt{2D}\,\xi(t) \ ,\\
y(t) & = vt.
\end{align}
\end{subequations}
In principle a constant force $f$ could also be included in the dynamics of $x$, but this is omitted here as it has no meaningful influence on the calculation. The distribution for $x(t)$ can be solved exactly~\cite{mazonka1999exactly}, giving in the long-time limit a Gaussian density 
\begin{equation}
    x(t) \sim \mathcal{N}\left(vt - v\tauR, \, \sigma^2\right) \ ,
\end{equation}
where $\tauR = \gamma/\kappa$ and $\sigma^2 = D\tauR$, as before. The heat flow can likewise be calculated exactly in the long-time limit, yielding $\dot{Q}_X = -\gamma v^2$.

The mean squared displacement during a time interval $\Delta t$ can be calculated exactly using (\ref{eq:MSD}) to give
\begin{align}
    \!\langle \Delta x^2 \rangle_{\mathrm{neq}} &= \sigma^2\!\left[2\left(1\!-\!e^{-\Delta t/\tauR}\right) \! - \! \beta\dot Q_X\tauR \left(1\!-\!e^{-\Delta t/\tauR} \right)^2 \right], \\
    &= 2\sigma^2\left(1\!-\!e^{-\Delta t/\tauR}\right) \! + \! \tauR^2 v^2 \left(1\!-\!e^{-\Delta t/\tauR} \right)^2 ,
    \end{align}
where we have used $\dot{Q}_X = -\gamma v^2$ in the second line. This then allows us to compute the MSD ratio directly for nonequilibrium dynamics with $v>0$, and equilibrium dynamics with $v=0$:
\begin{subequations}
\begin{align}
\frac{\langle\Delta x^2\rangle_\mathrm{neq}}{\langle\Delta x^2\rangle_\mathrm{eq}}
& = 1 + \frac{v^2 \tauR^2}{2 \sigma^2}\left(1-e^{-\Delta t/\tauR}\right)\\
& = 1 + \frac{v^2 \tauR}{2\sigma^2} \Delta t - \frac{v^2}{4\sigma^2} \Delta t^2 + \mathcal{O}\left(\Delta t^3\right)\ .
\end{align}
\end{subequations}
Inserting this into the heat estimator~\eqref{eq:heatestimator} gives
\begin{equation}
\beta \widehat{\dot{Q}_X} = -\frac{v^2 \tauR}{\sigma^2} + \frac{v^2}{2\sigma^2} \Delta t + \mathcal{O}\left(\Delta t^2\right)
\end{equation}
so that 
\begin{equation}
    \widehat{\dot{Q}_X} = - \gamma v^2 + \frac{\gamma v^2}{2\tauR} \Delta t + \mathcal{O}\left(\Delta t^2 \right) \ .
\end{equation}
The bias of the estimator is thus
\begin{subequations}
\begin{align}
\widehat{\dot{Q}_X} - \dot{Q}_X & = \frac{\gamma v^2}{2\tauR} \Delta t + \mathcal{O}\left(\Delta t^2\right) \\
& = -\frac{1}{2\tauR}\dot{Q}_X \Delta t + \mathcal{O}\left(\Delta t^2\right)\ ,
\end{align}
\end{subequations}
consistent with our prediction~\eqref{SI:bias}. 

We can also exactly compute the standard deviation of the estimator, giving
\begin{equation}
\mathrm{Std}\left(\beta \widehat{\dot{Q}_X}\right) = \frac{2}{\sqrt{N} \Delta t}\cdot\frac{1}{\left\langle \Delta x^2\right\rangle_\mathrm{eq}} \left[ \left(\frac{\left\langle \Delta x^2\right\rangle_\mathrm{neq}}{\left\langle \Delta x^2\right\rangle_\mathrm{eq}}\right)^2\left(\left\langle \Delta x^4\right\rangle_\mathrm{eq}-\left\langle\Delta x^2\right\rangle^2_\mathrm{eq}\right) + \left(\left\langle \Delta x^4\right\rangle_\mathrm{neq}-\left\langle\Delta x^2\right\rangle^2_\mathrm{neq}\right)\right]^{1/2}.
\end{equation}
In this case, both the nonequilibrium and equilibrium displacements are Gaussian-distributed, so this simplifies to
\begin{equation}
\mathrm{Std}\left(\beta\widehat{\dot{Q}_X}\right) = \frac{4}{\sqrt{N} \Delta t}\cdot\frac{\left\langle \Delta x^2\right\rangle_\mathrm{neq}}{\left\langle \Delta x^2\right\rangle_\mathrm{eq}}.
\end{equation}
Expanding this around $\Delta t=0$, we obtain
\begin{equation}
\mathrm{Std}\left(\beta \widehat{\dot{Q}_X}\right) = \frac{4}{\sqrt{N} \Delta t} - \frac{2}{\sqrt{N}}\beta \dot{Q}_X + \mathcal{O}\left(\frac{\Delta t}{\sqrt{N}}\right).
\end{equation}
As expected, here the leading term in $\Delta t^{-1}$ is consistent with our prediction~\eqref{eq:estimatorvariance}.

\subsection{Continuous information engine}
We next consider a ratcheting information engine rectifying fluctuations to raise a diffusive particle against gravitational force with magnitude $m g$. Here the trap and particle have respective positions $\lambda(t)$ and $x(t)$. The dynamics are
\begin{subequations}
\begin{align}
\lambda(t+t_s) & = \lambda(t) + \alpha\cdot(x-\lambda)\cdot\Theta(x-\lambda),\\
\dot{x} & = \frac{\kappa}{\gamma}(\lambda - x) -\frac{mg}{\gamma} + \sqrt{2D}\,\xi(t)\ .
\end{align}
\end{subequations}
We assume that the bead position is sampled at time intervals of $t_s$, and take $\alpha=2$ to ensure the trap does no work on the particle in the $t_s \rightarrow 0$ limit. For $\Delta t$ sufficiently small that the trap is approximately stationary, the particle step $\Delta x$ follows the conditional distribution
\begin{equation}\label{SI:infoengine:pDxgxl}
p(\Delta x\,|\,x,\lambda) = \mathcal{N}\left[\left(\lambda-x-mg/\kappa\right)\left(1-e^{-\Delta t/\tauR} \right), \, \sigma^2\!\left( 1-e^{-2\Delta t/\tauR} \right) \right].
\end{equation}
Here, as before, $\tauR = \gamma/\kappa$ and $\sigma^2 = D\tauR$. Following the details in the Appendix of Ref.~\cite{saha2021maximizing}, in the limit of infinite sampling frequency ($t_s\to 0$) the distribution of $u=\lambda-x$ is
\begin{equation}\label{SI:infoengine:pidef}
p(u) = \frac{1}{\sqrt{2\pi}\sigma \, \Phi(\delta_g)} \exp\left\{-\frac{(u-mg/\kappa)^2}{2\sigma^2}\right\}\Theta(u),
\end{equation}
where $\delta_g = mg/\kappa \sigma$ is the scaled effective mass and $\Phi(x)$ is the cumulative distribution function of a standard Gaussian,
\begin{equation}
    \Phi(x) \equiv \frac{1}{\sqrt{2\pi}}\int_{-\infty}^x e^{-y^2/2} dy.
\end{equation}
We then use this to compute the MSD, as
\begin{equation}
\begin{aligned}
\langle \Delta x^2\rangle_\mathrm{neq} & = \sigma^2\left(1 - e^{-2\Delta t/\tauR}\right) + \left(1-e^{-\Delta t/\tauR}\right)^2 \left\langle (\lambda-x-mg/\kappa)^2\right\rangle_{p(\lambda-x)}\\
& = 2\sigma^2\left(1-e^{-\Delta t/\tauR}\right) - \sigma^2  \delta_g \left(1-e^{-\Delta t/\tauR}\right)^2 \, v(\delta_g) \,,
\end{aligned}
\end{equation}
for engine scaled velocity~\cite{saha2021maximizing} 
\begin{equation}
v(\delta_g)\equiv\sqrt{\frac{2}{\pi}}\, e^{-\delta_g^2/2}\left[1 + \mathrm{erf}\left(\delta_g/\sqrt{2}\right)\right]^{-1}.
\end{equation}
As for the conventional engine, we have $\langle\Delta x^2\rangle_\mathrm{eq} = 2 \sigma^2\left(1-e^{-\Delta t/\tauR}\right)$, so the heat estimator expanded around $\Delta t=0$ gives
\begin{equation}
\begin{aligned}
\beta \widehat{\dot{Q}_X} & = \frac{2}{\Delta t}\left(1-\frac{\langle\Delta x^2\rangle_\mathrm{neq}}{\langle\Delta x^2\rangle_\mathrm{eq}}\right)\\
& = \frac{\delta_g \, v(\delta_g)}{\tauR} - \frac{\delta_g \, v(\delta_g)}{2 \tauR^2}\Delta t + \mathcal{O}(\Delta t^2).
\end{aligned}
\end{equation}
The distribution $p(u)$ also allows us to directly evaluate the true heat flow, yielding $\dot{Q}_X = \delta_g \, v(\delta_g)/\beta \tauR = \delta_g v(\delta_g) k_{\mathrm{B}}T/\tauR$, which agrees with the power output calculated in Ref.~\cite{saha2021maximizing}. Thus the bias of the estimator agrees to first order in $\Delta t$ with our general calculation:
\begin{equation}
\widehat{\dot{Q}_X} - \dot{Q}_X= -\tfrac{1}{2\tauR}\dot{Q}_X \Delta t + \mathcal{O}\left(\Delta t^2\right).
\end{equation}
We can also exactly compute the standard deviation of the estimator, giving
\begin{equation}
\mathrm{Std}\left(\beta \widehat{\dot{Q}_X}\right) = \frac{2}{\sqrt{N} \Delta t}\cdot\frac{1}{\left\langle \Delta x^2\right\rangle_\mathrm{eq}} \left[ \left(\frac{\left\langle \Delta x^2\right\rangle_\mathrm{neq}}{\left\langle \Delta x^2\right\rangle_\mathrm{eq}}\right)^2\left(\left\langle \Delta x^4\right\rangle_\mathrm{eq}-\left\langle\Delta x^2\right\rangle^2_\mathrm{eq}\right) + \left(\left\langle \Delta x^4\right\rangle_\mathrm{neq}-\left\langle\Delta x^2\right\rangle^2_\mathrm{neq}\right)\right]^{1/2}.
\end{equation}
Since in this case the equilibrium displacement $\Delta x$ is Gaussian-distributed, this simplifies to
\begin{equation}
\mathrm{Std}\left(\beta \widehat{\dot{Q}_X}\right) = \frac{2}{\sqrt{N} \Delta t}\cdot\frac{1}{\left\langle \Delta x^2\right\rangle_\mathrm{eq}} \left[ \left\langle \Delta x^2\right\rangle_\mathrm{neq}^2 + \left\langle \Delta x^4\right\rangle_\mathrm{neq}\right]^{1/2}.
\end{equation}
Evaluating this using $p(\Delta x \, |  \, x,\lambda)$~\eqref{SI:infoengine:pDxgxl} together with $p(\lambda-x)$~\eqref{SI:infoengine:pidef}, and expanding around $\Delta t=0$, gives
\begin{equation}
\mathrm{Std}\left(\beta \widehat{\dot{Q}_X}\right) = \frac{4}{\sqrt{N} \Delta t} - \frac{2}{\sqrt{N}}\beta \dot{Q}_X + \mathcal{O}\left(\frac{\Delta t}{\sqrt{N}}\right).
\end{equation}
Again, here the leading term in $\Delta t^{-1}$ is consistent with our general prediction~\eqref{eq:estimatorvariance}.

\section{Details of kinesin motor simulations}

\subsection{Discrete model for kinesin dynamics}
\label{SI:kinesinmodel}
\begin{figure}[h]
    \centering
    \includegraphics[width =0.8\linewidth]{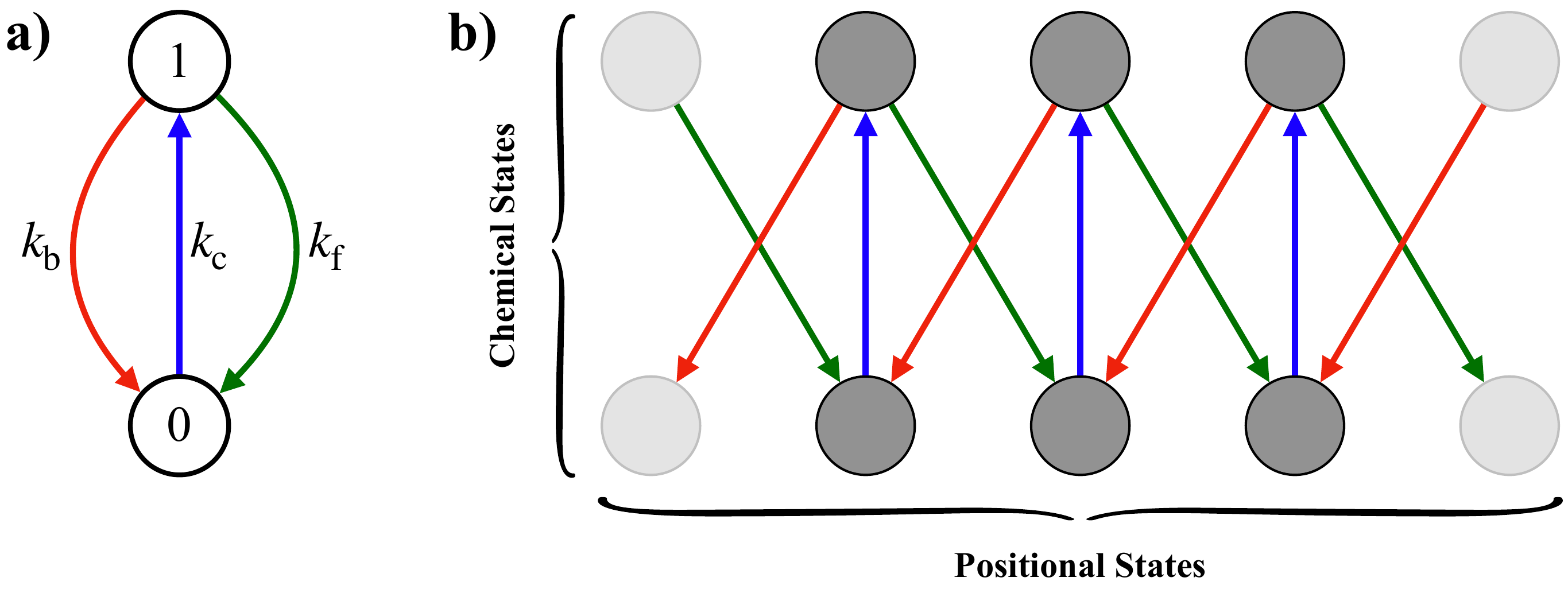}
    \caption{Diagram illustrating the discrete-state stochastic model for kinesin dynamics. a) Chemical states, labeled $0$ and $1$, and the three different transitions between them. There are two cycles (blue transition then red transition, or blue transition then green transition) respectively corresponding to forward and backward steps. b) Chemical states ``unwrapped" over the (horizontal) space of mechanical states.}
    \label{fig:kinesinmodeldiagram}
\end{figure}

To model the dynamics of a kinesin motor in the main-text section \emph{Kinesin pulling diffusive cargo}, we use the two-state discrete model from \cite{ariga2018nonequilibrium}, which was parameterized by fitting to experimental force-velocity curves. This model, while fit to dynamics with only equilibrium noise, has previously been shown to reproduce the experimentally observed velocity increase in the presence of nonequilibrium noise~\cite{ariga2021noise}. As shown in Fig.~\ref{fig:kinesinmodeldiagram}, the model features two chemical states with two cycles corresponding respectively to forward or backward steps. The forward ($k_\mathrm{f}$) and reverse ($k_\mathrm{b}$) step rates depend on the cargo-position-dependent force from the linker according to
\begin{subequations}
\begin{align}
k_\mathrm{f} & = k_\mathrm{f}^0 \exp\left[\beta d_\mathrm{f} \kappa(x-y)\right],\\
k_\mathrm{b} & = k_\mathrm{b}^0 \exp\left[\beta d_\mathrm{b} \kappa(x-y)\right]\ .
\end{align}
\end{subequations}
Here $k_\mathrm{f}^0$ and $k_\mathrm{b}^0$ are bare rate constants, and the parameters $d_\mathrm{f}$ and $d_\mathrm{b}$ quantify the respective dependences of the forward and reverse rates on the linker force $\kappa(x-y)$ on the motor. Forward or reverse stepping must be preceded by an activation step which occurs with rate $k_\mathrm{c}$, which is independent of the linker force.

\label{SI:modeldiagram}

Following Ref.~\cite{ariga2018nonequilibrium}, we take the parameter values $k_\mathrm{f}^0 = 1002\,$s$^{-1}$, $k_\mathrm{b}^0 = 27.9\,$s$^{-1}$, $k_\mathrm{c} = 102\,$s$^{-1}$, $d_\mathrm{f} = 3.61\,$nm, and $d_\mathrm{b}=1.14\,$nm. We numerically simulate the stochastic dynamics of the coupled motor-cargo system in Python with a simulation timestep of $10^{-5}\,$s.

\subsection{Additional simulation results}
\begin{figure}[h]
    \centering
    \includegraphics[width =0.5\linewidth]{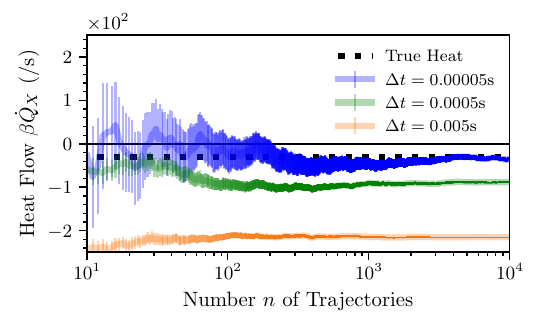}
    \caption{Inferring heat flows for a kinesin motor pulling a diffusive cargo. Mean and standard error (error bars) for the heat estimator, as a function of the number $n$ of trajectories of duration $1$\,s, for different values of the sampling time $\Delta t$. Black dotted line indicates the true heat flow calculated from simulation data of full $x$ and $y$ trajectories. Simulation parameters are $f=0$ and $D_\mathrm{eq} = 126,680$\, nm$^2$/s.}
    \label{fig:fig4}
\end{figure}

Figure~\ref{fig:fig4} shows the estimator's performance as a function of the number $n$ of trajectories, for different values of the sampling time $\Delta t$, benchmarked against the true heat flow calculated from full knowledge of both cargo and motor dynamics. The estimator behaves as we calculated in App.~\ref{SI:bias} and \ref{SI:variance}, with bias decreasing with smaller $\Delta t$ (at large $n$), and variance decreasing with increasing $n$ and increasing $\Delta t$. For $\Delta t\approx 0.00005$ s, around a thousand trajectories suffice to estimate the heat flow. 

\label{SI:morekinesinresults}

\bibliography{main}